\documentclass[showkeys,showpacs,amssymb,eqsecnum,superscriptaddress]{revtex4}

\usepackage{dcolumn}
\usepackage{graphics}
\usepackage{latexsym,epsf}
\usepackage{amsfonts}
\usepackage{hyperref}
\begin{document} 

\title{Zeta Functions in Brane World Cosmology}

\author{Antonino Flachi}
\email[Email: ]{flachi@yukawa.kyoto-u.ac.jp}
\affiliation{Yukawa Institute for Theoretical Physics, Kyoto University, 
Kyoto 606-8502, Japan}
\author{Alan Knapman}
\email[Email: ]{a.h.knapman@ncl.ac.uk}
\affiliation{School of Mathematics and Statistics, University of  
Newcastle upon Tyne\\ 
Newcastle upon Tyne, NE1 7RU, UK}
\author{Wade Naylor}
\email[Email: ]{naylor@yukawa.kyoto-u.ac.jp}
\author{Misao Sasaki}
\email[Email: ]{misao@yukawa.kyoto-u.ac.jp}
\affiliation{Yukawa Institute for Theoretical Physics, Kyoto University, 
Kyoto 606-8502, Japan}

\begin{abstract}
We present a calculation of the zeta function and of the functional
determinant for a
Laplace-type differential operator, corresponding to a scalar field in a higher dimensional 
de Sitter brane background, which consists of
a higher dimensional anti-de Sitter bulk spacetime bounded by a de Sitter section, 
representing a brane.
Contrary to the existing examples, which all make use of conformal transformations, we
evaluate the zeta function working directly with the higher dimensional wave
operator. We also consider a generic mass term and coupling to curvature, generalizing previous results.
The massless, conformally coupled case is obtained as a limit of the general result and compared
with known calculations.
In the limit of large anti-de Sitter radius, the zeta determinant for the ball is recovered in perfect
agreement with known expressions, providing an interesting check of our result and an alternative
way of obtaining the ball determinant.

\hfill \vbox{ \hbox{YITP-04-59} }
\end{abstract}

\pacs{04.62.+v, 11.10.Kk, 02.30.-f}
\keywords{Quantum fields in curved spacetime, Extra dimensions, Zeta functions}

\maketitle
\today

\newcommand{\g}{g_{\mu\nu}}
\newcommand{\kk}{q}
\newcommand{\vp}{\varphi}
\newcommand{\remark}{{\bf ***}}
\newcommand{\mm}{{\bf ********}}
\newcommand{\nn}{\nonumber\\}
\newcommand{\hsi}{\hat{\sigma}}
\newcommand{\hrho}{\hat{\rho}}
\newcommand{\hl}{\hat{\lambda}_\k}
\newcommand{\si}{\sigma}
\newcommand{\mubar}{\bar{\mu}}
\newcommand{\dlt}{\bigtriangleup}
\newcommand{\beq}{\begin{equation}}
\newcommand{\eeq}{\end{equation}}
\newcommand{\bed}{\begin{displaymath}}
\newcommand{\eed}{\end{displaymath}}
\def\bea{\begin{eqnarray}}
\def\eea{\end{eqnarray}}

\section{Introduction}

The idea of living on a membrane embedded in a higher-dimensional spacetime has attracted
enormous attention over the past few years, the main motivation being the fact that the 
localization of particles on branes provides an alternative to the standard picture of
Kaluza-Klein compactification \cite{rubakov,akama}.

A popular example is the Randall-Sundrum model \cite{RS1}, which considers a five-dimensional
slice of Anti-de Sitter (AdS) spacetime with the extra dimension compactified to an orbifold having
two 3-branes of opposite tension located at its fixed points. This results in a
non-factorizable spacetime, which, as Randall and Sundrum have shown, has the important
consequence that the effective four-dimensional scale on the negative-tension brane turns out
to be exponentially suppressed relative to the higher-dimensional scale. This was originally
proposed as a solution to the hierarchy problem by explaining the very small ratio (large
hierarchy) of some 17 orders of magnitude between the electroweak scale and the Planck scale
observed on our brane, identified as that with negative tension. In this model, the hierarchy
arises as a geometrical effect, with gravity being strongly localized around the
positive-tension brane.

Pushing this idea further, Randall and Sundrum \cite{RS2} showed that the negative-tension
brane may actually be absent and the extra dimension possibly infinite in size. In that case,
we live on the positive-tension brane with gravity confined around it, recovering
its standard Einstein form from an effective four-dimensional point of view. This
result does not solve the hierarchy problem, but has stimulated the construction of many new
cosmological models. Here we concentrate our attention on the de Sitter brane model, 
relevant in the construction of brane world models of inflation. 

The problem of studying quantum effects in such scenarios is, then, naturally posed. It is
immediately clear that when a quantum field is considered on such background spacetimes,
quantum effects may play a significant role, the simple contribution that they give to the
brane and bulk cosmological constants being an evident example.

A number of people, inspired by previous work in Kaluza-Klein theories \cite{tomsplb,cw},
have investigated the possibility that quantum effects from bulk fields could play a role
in the consistency of such models by providing a sensible mechanism of stabilization in the
Randall-Sundrum two-brane model and in some higher-dimensional generalizations. 
Various authors have dealt with the problem of quantum effects in brane models and some references 
are~\cite{GR,T,BR,GPT,FT,FMT1,FMT2,SS,MN,KT}.
The lowest order quantum corrections arising from bulk fields on the Randall-Sundrum background
have been calculated in a variety of ways, and then extended to some general classes of
higher-dimensional spacetimes in Refs.~\cite{FGPT,FP,Fl}. 
Ref.~\cite{GP} considers the case of scalar and gauge fields in the Randall Sundrum model and
interprets the result in terms of the AdS/CFT correspondence, showing explicitly that quantum
effects could provide a sensible stabilization mechanism.

It is worth mentioning that in most of the previous work, the geometry of the branes was
assumed to be flat, which greatly simplifies the study of quantum effects. When the branes
are curved, as happens, for example, in de Sitter or hyperbolic brane models, and where the
Casimir energy could have some effect on the cosmological evolution of the brane, the
situation becomes more complicated.

Some work in this direction has been carried out in
Refs.~\cite{NOZ,NS,ENOO,MNSS}. Specifically, in Refs.~\cite{NOZ,NS}, the vacuum energy for a
massless conformally coupled scalar field in a brane world corresponding to de Sitter branes
in an AdS bulk has been evaluated. This calculation, which is technically the simplest, is
carried out by working in a conformally related spacetime, similar in form to the Einstein
universe.  Zeta function regularization is employed and the result shows that the vacuum
energy is zero for the one brane configuration. These results have been extended to the case
of conformally coupled Majorana spinors in Ref.~\cite{MNSS}, still by making use of conformal
transformations.
The cases of a massless conformally coupled scalar field and of a massive minimally coupled
field for de Sitter branes embedded in both de Sitter and AdS bulks have been
considered in Ref.~\cite{ENOO}, where the vacuum energy is computed once again by using
conformal and zeta regularization techniques and is found to be zero for the one brane case.
Refs.~\cite{NOZ,Norm} consider a somewhat related set-up and compute the effective
action for scalar fields in an AdS bulk bounded by AdS
branes, still by making use of conformal transformations.

Apart from the relatively simple case of massless, conformally coupled field, the general case
has not yet been studied.
The present paper is devoted to providing a new derivation of the zeta function and of the
functional determinant for a scalar field in a de Sitter brane model. The method is very
general and applies, in principle, to a number of situations, where the methods based on
conformal transformations or dimensional reduction cannot be applied at ease.
 
We perform the calculation by directly working in the higher-dimensional spacetime and
evaluate the zeta function for the original higher-dimensional wave operator. With respect
to the previously studied cases, we have to deal with two main problems. The first is to
compute a zeta function where the operator spectrum is not known, for which we adopt the
technique developed in Refs.~\cite{IGM,BKK,BKKM,KM,BEK,BKD,Klausbook}. This general technique,
elaborated in various forms, allows one to obtain the zeta function using only the knowledge
of the basis functions. Such a method is applicable whenever an implicit equation satisfied
by the eigenvalues is known, and has already been applied in the context of brane models. We
stress, however, that in the case of flat branes the basis functions can be conveniently
expressed in terms of Bessel functions, which greatly simplifies the calculation. The second
main problem is that, for the class of spacetimes we consider, namely an AdS bulk
bounded by a de Sitter brane, the basis functions are expressed as a combination of generalized
Legendre functions, which are considerably less manageable. However, the method developed in
Ref.~\cite{BKK} proves to be particularly useful and we closely follow their approach in our
calculation. 

The structure of the paper is as follows. In the next section, we discuss the spacetime
configuration of the de Sitter brane model and solve the higher-dimensional scalar wave
equation on such a background spacetime. For the sake of generality, we consider a massive
scalar field coupled to the higher-dimensional curvature. In the subsequent section, after
having described the general technique of Ref.~\cite{BKK}, we pass to the main task at hand,
which is the evaluation of the zeta function for a de Sitter brane model. 
After having obtained the general result, we consider two limiting cases, which provide a non
trivial check on the method used as well as the actual calculation.
First we specify the result to the conformally coupled case and compare with that of
Refs.~\cite{NOZ,NS,ENOO,MNSS}. Then we consider the limit of large AdS curvature
radius $\ell$, which should reproduce a ball-like geometry. The last section is left for
conclusions. Several technical results regarding the asymptotic expansion of the generalized
Legendre functions, as well as the derivations of certain results used in the calculation, are
reported in the appendices.

\section{Scalar fields on de Sitter brane backgrounds}
\label{sec2}

\noindent 

The background configuration we consider is described by the action
\beq
S = \int d^5x \sqrt{-g^{(5)}} \left( {1\over 2 \kappa^2}( {\cal R} - 2\Lambda_5)\right) -
\int d^4x \sqrt{-g^{(4)}} ~\tau~, 
\eeq
where 
\beq
\Lambda_5 = - {6\over \ell^2}
\eeq
is a higher-dimensional, negative, bulk cosmological constant term, $\tau$ is the brane
tension and $\kappa^2$ is proportional to the five dimensional gravitational constant. The
corresponding solution of the setup, with the usual cosmological symmetries, has an AdS
 five-dimensional geometry; $\ell$ is the radius of the five dimensional AdS space.

A convenient scaling of the time coordinate allows one to write the metric as
\beq
  ds^2 = dr^2+ (H\ell)^2 \sinh^2(r/\ell)[-dt^2+H^{-2}e^{2Ht}d{\bf x}^2_{3}]~, 
\label{uno}
\eeq
where the coordinate $r$ parameterizes the extra dimension and the time coordinate
corresponds to the cosmic time parameter on the brane. For notational convenience, we define
\beq
  a(r)= H \ell \sinh(r/\ell).
\label{scalefact}
\eeq

A ${\mathbb Z}_2$-symmetric brane world can then be constructed in a standard way by taking
two slices of the space and gluing them along the brane located at $r_0$. In such a case, the
junction conditions at the brane give the Friedmann equation
\beq
H^2 = {\Lambda_5 \over 6} + \left({\kappa^2\over 6}\right)^2 \tau^2~.
\eeq
In the present case, the Hubble parameter, $H$, is constant, so that the brane geometry is de
Sitter. The Hubble parameter is related to the brane position $r_0$ according to
\beq
H^2 = {1\over \ell^2 \sinh^2(r_0/\ell)}~.
\eeq

On such a background, we consider a bulk scalar field and ignore its back reaction.
It satisfies the Klein-Gordon equation
\beq
(-\Box_{\rm E}+m^2+\xi {\cal R})~\varphi(x,r) = 0~. \label{eomKG}
\eeq
For the sake of generality, we consider the brane to be $D$- rather than four-dimensional,
for the remainder of the present section. Furthermore, for later convenience, we transform to Riemannian signature.
Hence, $\Box_{\rm E}$ is the $(D+1)$-dimensional d'Alembertian on Riemannian space and ${\cal R}$ is the scalar
curvature, given by
\bea
  {\cal R} &=& a^{-2}(r)~{\cal R}_\Sigma -\left[2D {a''(r)\over
a(r)}+D(D-1) \left({a'(r)\over a(r)}\right)^2\right]~,
\label{scal}
\eea
where ${\cal R}_\Sigma=D(D-1)H^2$ is the scalar curvature of the de Sitter brane, which is a $D-$sphere of radius $H^{-1}$.

We are interested in finding the eigenmodes $\varphi_{n,j}(x,r)$ and eigenvalues
$\lambda_{n,j}^2$ of the above field operator, defined by
\beq
(-\Box_{\rm E}+m^2+\xi {\cal R})~\varphi_{n,j}(x,r) = {\lambda_{n,j}^2}
~\varphi_{n,j}(x,r)~. \label{eom}
\eeq
Let us assume that the modes are separable in the variables $x$ and $r$:
\beq
  \varphi_{n,j}(x,r)=\phi_{j}(x) f_{n,j}(r),
\eeq
where the spherical eigenfunctions satisfy 
\beq
  -\Box_\Sigma\phi_{j}(x)=j\left(j+ D-1\right) H^2 \phi_{j}(x),
\label{sph}
\eeq
where $\Box_\Sigma$ is the d'Alembertian on the de Sitter section and $d_j$ the degeneracy factor, 
\beq
  d_j=(2j+D-1){(j+D-2)!\over j!(D-1)!}
\label{degn}
\eeq
with $j=0,1,2,\dots$. Using Eq.~(\ref{sph}) in Eq.~(\ref{eom}) allows
us to find the
equation of motion for the radial eigenfunctions: 
\beq
  \left[-a^{-D}(\partial_r a^D \partial_r )+a^{-2}j\left(j+ D-1\right)H^2+m^2+\xi{\cal
R}\right] f_{n,j}(r)=\lambda_{n,j}^2 ~f_{n,j}(r)~, \label{radial}
\eeq
whose solution can be written in terms of toroidal Legendre functions
\beq
 f_{n,j}(\eta) = (H\ell)^{1/2-D/2} \sinh^{1/2-D/2} (\eta) \left[A_{n,j} P_{i
\omega_{n,j}-1/2}^{-\nu_j} (\cosh \eta ) 
+ B_{n,j} Q_{i \omega_{n,j}-1/2}^{-\nu_j} (\cosh \eta)\right]~,
\label{modes}
\eeq
where we have defined $\eta=r/\ell$. The order and degree of the associated
Legendre functions are
\beq
  \nu_j= \left(j+ {D-1\over 2}\right)\qquad\mbox{and}\qquad\omega_{n,j}
=\sqrt{\ell^2\lambda^2_{n,j}-\sigma^2}~, 
\label{eigen}
\eeq
where 
\beq
  \sigma^2 = \ell^2 m^2+\xi\ell^2{\cal R}+D^2/4~.
\eeq 
Regularity at the origin implies that $B_{n,j}=0$, as follows from examining the
small-argument behavior of the generalized Legendre functions.

Thus, our eigenfunctions take the form
\beq
  f_{n,j}(\eta) = A_{n,j} a^{\frac {1-D} 2}(\eta)~P_{i \omega_{n,j}-1/2}^ {-\nu_j} (\cosh \eta ), \label{numod}
\eeq
with $a(\eta)= H \ell \sinh\eta$.

As well as the solution in the bulk, we must also consider the boundary condition on the
brane, which can be obtained by integrating (\ref{radial}) across the brane.
In general, the ${\mathbb Z}_2$ symmetry allows us to choose either an untwisted field
configuration, such that $f(\eta)=f(-\eta)$, corresponding to Robin boundary conditions
\beq
\partial_\eta f_{n,j}(\eta_0)=-{2D\xi \over \ell} \coth(\eta_0) f_{n,j}(\eta_0)~,
\eeq
or alternatively a twisted field configuration, such that $f(\eta)=-f(-\eta)$,
corresponding to Dirichlet boundary conditions
\beq
  f_{n,j}(\eta_0) = 0~.
\label{eigeq}
\eeq
In the remainder of the paper we focus our attention to Dirichlet boundary conditions.

\section{Zeta function for de Sitter brane models}
\label{sec3}

\subsection{General method}
\label{subsec3}
The main scope of this paper is to compute the zeta function and the functional determinant for a
bulk scalar field on a de Sitter brane background. For calculational simplicity, we take a field
obeying Dirichlet boundary conditions, although the method can be applied to more general
situations with few calculational modifications. To deal with the case of de Sitter brane
models, we follow the approach of Refs.~\cite{IGM,BKK,BEK,BKD,Klausbook}, where a calculational
technique for $\zeta$-functions of differential operators on manifolds with boundaries with
explicitly unknown spectra has been developed and applied (see Refs.~\cite{BKK,BKKM,KM}) to
evaluate the one-loop contribution from the graviton, and matter fields, to the Hartle-Hawking
wave function. Here we summarize their method.

It is a well-known fact that the one-loop effective action can be written as
\beq
\Gamma^{(1)} = {1\over 2} ~\mbox {Tr} \ln {\Delta}~,
\eeq
where $\Delta$ is a second-order differential operator, in our case given by
\beq
  \Delta = -\Box_E + m^2 + \xi {\cal R}~. \label{3.1}
\eeq
$\Gamma^{(1)}$ can be expressed in terms of a generalized $\zeta$-function
\beq
\zeta(s) = \sum_{\lambda} \lambda^{-s}~,
\eeq
with $\lambda$ being the eigenvalues of the operator $\Delta$, which we assume to be positive definite. One has
\beq
\Gamma^{(1)} = -{1\over 2} \zeta'(0) - {1\over 2} \zeta(0) \ln \mu^2~,
\label{effact}
\eeq
where $\mu$ is the renormalization scale. Thus, we see that the main problem is reduced to
that of evaluating the $\zeta$-function and its derivative at $s=0$.

Usually, the computation of the $\zeta$-function requires explicit knowledge of the
spectrum. However, in many situations of interest, the eigenvalues are not explicitly
known. To bypass this kind of problem, various authors have developed a calculational
technique that allows one to evaluate the $\zeta$-function and related quantities like
functional determinants, Casimir energies and effective actions, when such information on the
eigenvalues is lacking and the only knowledge of the spectrum is via an implicit equation
\cite{BKK,BEK,BKD,Klausbook}. Generally, one considers the eigenvalue problem associated with
the operator $\Delta$:
\beq
\Delta u_\nu(x) = \lambda u_\nu(x)~,
\label{g0}
\eeq
where the parameter $\nu$ enumerates the independent solutions of (\ref{g0}). A degeneracy factor $d(\nu)$ is associated with each
$\nu$. Imposing the relevant boundary conditions leads to an equation of the form
\beq
F(\lambda,\nu)=0~,
\label{g1}
\eeq
where the function $F(\lambda,\nu)$ depends on the mode functions, the eigenvalues $\lambda$,
the index $\nu$ and eventually other parameters inessential for the present discussion.

To evaluate the $\zeta$-function it is not necessary to solve the previous equation, as is
clear by making use of the residue theorem, which allows one to write
\beq
\zeta(s) = {1\over 2\pi i} \int_{\cal C} \omega^{-s} {d\over d\omega} \sum_\nu d(\nu)\ln F(\omega,\nu) d\omega~,
\label{g2}
\eeq
where the contour $\cal C$ is chosen to enclose all the positive solutions of (\ref{g1}) in the complex $\omega-$plane.

For the explicit calculation, it is convenient to express this contour integral as an
integral along the real line, which can be achieved by appropriately deforming the contour
${\cal C}$. Typically, if the function $F(\omega,\nu)$ satisfies certain properties, as in the case we will consider in the next section, a 
choice of a contour like the one in Fig.~\ref{figcont}, allows one to rewrite the integral (\ref{g2}) as
\beq
\zeta(s) = {\sin\pi s\over \pi} \int_0^\infty z^{-s} {d\over dz}  \sum_\nu d(\nu) \ln F(z,\nu) dz~.
\label{g3}
\eeq
The source of divergences in the above expression come from the large $\nu$ behavior and the
integration over $z$. We need to check that large values of $s$ indeed regulate
$\zeta(s)$. Let us first presuppose that this is possible and, assuming $s$ to be large
enough, we swap the order of integration and summation and consider the asymptotic behavior
of the integral
\beq
\int_0^\infty z^{-s} {d\over dz}\ln F(z,\nu) dz~ 
\label{g4}
\eeq
for $\nu\rightarrow\infty$ and its integrand for $z\rightarrow\infty$ and
$z\rightarrow0$. Now, it is possible to obtain a uniform asymptotic expansion
$\Phi(\nu,z/\nu)$ of $\ln F(z,\nu)$ such that $\nu\rightarrow\infty$, while the
ratio $z/\nu$ is held fixed. Thus
\beq
\ln F(z,\nu) \sim \Phi(\nu,z/\nu)~,
\label{g5}
\eeq
for large $\nu$. There are two important features of this expansion, aside from its
uniformity: it has a power-law behavior of fixed order in $\nu$, as we shall illustrate for
our case, and it is valid in the full range of the ratio $0\leq z/\nu<\infty$. If the
integral (\ref{g4}) is finite for a large enough value of $s$, which is true in our case,
then uniformity, together with these two properties ensures the convergence of the sum over
(\ref{g4}), for sufficiently large $s$. This can be proved under quite general assumptions, but
here we content ourselves with showing that this is true for the case under study. Finally, a
simple rescaling allows us to write
\beq
\zeta(s) = {\sin\pi s\over \pi} \sum_\nu d(\nu) \nu^{-s} \int_0^\infty z^{-s} {d\over
dz}\Phi(z,\nu)dz~.
\label{g6}
\eeq
This verifies that it is possible to regularize the divergences by a suitable choice of
$s$. We now interchange the integration and summation back to their original order and write
\beq
\zeta(s) = {\sin\pi s\over \pi} \int_0^\infty z^{-s} {d\over dz}I(z,s)dz ~,
\label{g7}
\eeq
where $I(z,s)$ is given by 
\beq
I(z,s) = \sum_\nu d(\nu) \nu^{-s} {d\over dz}\ln F(z,\nu) ~.
\label{g8}
\eeq

In order to further proceed with the evaluation of the $\zeta$-function, we must analytically
continue to $s=0$. We expand the sum (\ref{g8}) around small values of $s$, which generally
develops a pole:
\beq
I(z,s) \sim {I^{\rm pole}(z) \over s} + I^{\rm R}(z) + O(s)~.
\label{fru}
\eeq
It is now possible to use the following lemma along with the properties of the asymptotic
expansion to compute $\zeta(s)$. Consider a function ${\it f}(x)$ which is analytic at
$x=\epsilon$, for some small $\epsilon$, and has the following general asymptotic behavior,
when $x\rightarrow\infty$, 
\beq
{\it f}(x)=\sum_{k=1}^{\rho_k<N}\left({\it f}_k+\bar{\it f}_k\ln x\right)x^{\rho_k}+[{\it
f}]_{\rm log} \ln x+ [{\it f}]_{\rm reg}+O(x^{-1})~,\qquad\rho_k>0~,
\label{phix}
\eeq
where the subscripts ${\rm log}$ and ${\rm reg}$ refer to the solely logarithmic and regular
(non-singular) parts of ${\it f}(x)$ in the large $x$ limit. Then, there exists the analytic
continuation of the integral
\beq
\int_\epsilon^\infty {dx\over x^s} {d\over dx} {\it f}(x)={[{\it f}]_{\rm log}\over s}+
[{\it f}]^{\infty}_{\epsilon} 
+O(s)~,
\label{continue}
\eeq
where $[{\it f}]^{\infty}_{\epsilon} \equiv [{\it f}]_{\rm reg}-{\it f}(\epsilon)$, 
for example see Ref.~\cite{BKK}. 

On the basis of the uniform asymptotic expansion of the eigenfunctions, it is possible to
prove that $I^{\rm pole}(z)$ and $I^{\rm R}(z)$ behave as (\ref{phix}) and we also assume
that $[I^{\rm pole}]_{\rm log}=0$, which we shall show to be true in the case of interest to
us. It is now a simple matter to apply the result (\ref{continue}) to $I^{\rm pole}(z)$ and
$I^{\rm R}(z)$ to get
\beq
\zeta(s) = [I^{\rm R}]_{\rm log}+[I^{\rm pole}]^\infty_0
+s\left\{ [I^{\rm R}]^\infty_0-\int^\infty_0 dz \ln z^2 {d I^{\rm pole}(z)\over dz} \right\}
+O(s^2)~,\qquad s\rightarrow 0~.
\label{lemma}
\eeq
This equation can then be used to get the value of the $\zeta$-function and its derivative at
$s=0$.

\subsection{Explicit evaluation of the $\zeta$-function}


\begin{figure}
\scalebox{0.6} {\includegraphics{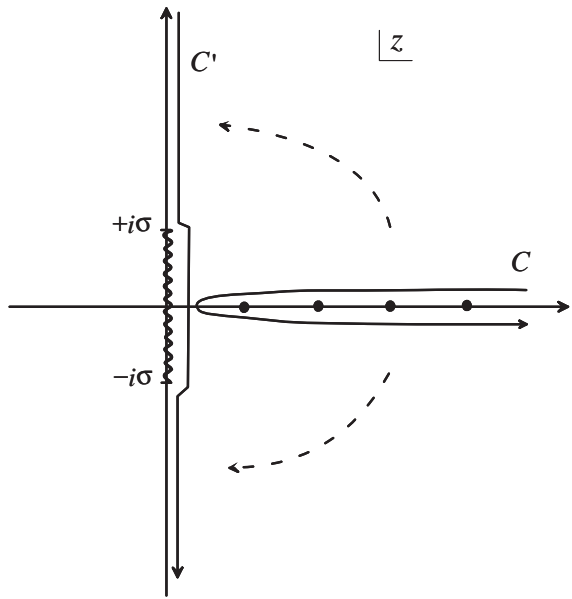} }
\caption{ 
The original contour (enclosing the positive real zeros) and
the deformed contour lying along the imaginary axis. The deformed contour $C'$ is closed at infinity by a semi-circle
of infinite radius.
\label{figcont}} 
\end{figure}


We now pass to the explicit evaluation of the $\zeta$-function for the scalar field on
the de Sitter background, described in section (\ref{sec2}).
We have seen that, for Dirichlet boundary conditions, the eigenvalues are given by the
solution of the implicit equation
\beq
F(\omega_{n,j};\nu_j) \equiv 2^{\nu_j}\Gamma(1+\nu_j) 
P_{i \omega_{n,j}-1/2}^ {-\nu_j} (\cosh\eta_0)=0~.
\label{eigeqn}
\eeq
Note that we have multiplied the Legendre function by the factor
$2^{\nu_j}\Gamma(1+\nu_j) $ as this does not change the solution of the eigenvalue
equation (\ref{eigeq}) and, on the other hand, produces some simplifications in the
calculations at a later stage. It is also clear that any factor independent of $z$ does
not affect the contour integral (\ref{g2}).

Thus the $\zeta$-function can be expressed by the double sum
\beq
  \zeta(s) = \sum_{n,j} d_j ~\lambda_{n,j}^{-2s} = \ell^{2s} \sum_{n,j}d_j ~\big(
\omega^2_{n,j}+\sigma^2\big)^{-s}, \label{zeta}
\eeq
with $\omega_{n,j}$ defined in (\ref{eigen}) being the solutions to (\ref{eigeqn}). 
As described in the previous subsection, we can use the residue theorem 
to express the $\zeta$-function as a contour integral, in terms of the complex parameter $\omega$
\beq
  \zeta(s) = {\ell^{2s} \over2\pi i}\sum_j d_j \int_{\cal C}{d\omega \over~(\omega^{2}+\sigma^2)^s}
{d\over d\omega}~ \ln F(\omega,\nu_j)~, 
\label{cont}
\eeq
where the contour ${\cal C}$ is chosen to enclose the real positive zeros of $F(\omega,\nu_j)$.

By appropriately deforming the contour (see Fig.~\ref{figcont}) and by performing some formal
manipulations, we arrive at
\beq
\zeta(s) ={\sin\pi s\over \pi }\int^\infty_{0}{dz^2 \over z^{2s}}{d\over d z^2}~I(z^2,s)~,
\label{intint}
\eeq
where
\beq
  I(z^2,s)=\sum_j d_j {1\over\nu_j^{2s}}~
           \ln[~2^{\nu_j}\Gamma(1+\nu_j)~P^{-\nu_j}_{\sqrt{\nu_j^2 z^2 \ell^2 + \sigma^2}
-\frac 1 2}(\cosh\eta_0)]~. 
\label{im}
\eeq
As in the previous subsection, we split the contributions to the $\zeta-$function into one
regular plus one polar piece:
\beq
\zeta(s)= \zeta_R(s) + \zeta_P(s)~,
\eeq
where
\beq
\zeta_R(s) = 
{\sin \pi s \over \pi} \int_0^\infty dz^2 z^{-2s}{d\over dz^2}I^R(z^2,s)~,
\eeq
and
\beq
\zeta_P(s) = 
{\sin \pi s \over \pi} \int_0^\infty dz^2 z^{-2s}{d\over dz^2} I^{\rm pole}(z^2,s)~.
\eeq
In the previous two expressions $R$ means that we have to take the regular part of the
large$-z$ expansion of the integrand, whereas $pole$ refers to the pole part at large $z$.

The integrand functions have the asymptotic behavior (\ref{phix}), as shown in Appendix
\ref{uniasy}; we are, therefore, justified in applying the lemma discussed previously. 
Then, it is easy to see that
\beq
\zeta_R(s) = [I^R]_{\rm log} + s([I^R]_{\rm reg} - I^R(0)) + O(s)~,
\eeq
and 
\beq
\zeta_P(s) = [I^{\rm pole}]_{\rm log} -s \int_0^\infty dz^2 \ln z^2 {dI^{\rm pole}(z^2)\over dz^2} +
O(s)~.
\eeq
The derivative can now be calculated easily and the previous results combine to give
\beq
  \zeta(0)= [I^{\rm R}]_{\rm log}+[I^{\rm pole}]_{\rm reg} - I^{\rm pole}(0)~,
\label{zetz}
\eeq
and
\beq
  \zeta'(0) = [I^{\rm R}]_{\rm reg} - I^{\rm R}(0) - \int^\infty_0 dz^2 \ln z^2 
{d I^{\rm pole}(z^2)\over dz^2}~,
\label{zetp}
\eeq
where we have anticipated the fact that $[I^{\rm pole}]_{\rm log}=0$, as shown in
Appendix \ref{appprop}. The asymptotic expansion of $I(z^2,s)$ is also given in Appendix
\ref{appprop}, where the various pieces appearing in (\ref{zetz}) and (\ref{zetp}) are
obtained.

For $\zeta(0)$, $[I^{\rm R}]_{\rm log}$, $[I^{\rm pole}]_{\rm reg}$ and $I^{\rm pole}(0)$ 
are needed, and Eqs.~(\ref{Ilog}), (\ref{ipreg}) and (\ref{ipzu}) provide them. 
Some algebra leads to the desired result:
\beq
  \zeta(0) = {1\over6}{17\over1920}+\frac{\sinh^2\eta_0}{32}
\left(1+2\sinh^2\eta_0\right) \left(\sigma^2-1/4\right)
- \frac{\,{\sinh^4 \eta_0 }}{48}\left(\sigma^2-1/4\right)^2~.
\label{reszetazero}
\eeq

On the other hand, the evaluation of $\zeta'(0)$ requires the knowledge of 
$[I^{\rm R}]_{\rm reg}$, $I^{\rm R}(0)$ and of the integral piece
appearing in Eq.~(\ref{zetp}). These are calculated in 
Appendix \ref{appprop} and the results are reported in formulas (\ref{logM}), (\ref{iru})
and (\ref{IReg}), which, combined together, lead to the following
expression for $\zeta'(0)$:
\bea
\zeta'(0)=
\zeta(0)\ln (\ell \sinh\eta_0)^2 + \Delta_1 + \Delta_2 + \Delta_3 + \Delta_4
\label{reszetaprimezero}
\eea
where
\beq
\Delta_1 \equiv {47\over 9216}-{1\over 64}\zeta'(0,1/2)+{1\over 24} \zeta'(-1,1/2)
+{1\over24} \zeta'(-2,1/2)-{1\over 6} \zeta'(-3,1/2)+{1\over 12} \zeta'(-4,1/2)
\label{zetap5}
\eeq
\beq
\Delta_2 \equiv
-{137\over7200}\left.{d^5\phi\over dx^5}\right|_{x=0}  
+{1\over120}\int_0^\infty dx\ln x{d^6\phi\over dx^6}+\int^{\infty}_{2/3} x^{-6} \phi(x)~dx~, 
\eeq
\beq
\Delta_3 \equiv
-~i\int_0^\infty dx{(ix+3/2)^3/3-(ix+3/2)/12\over e^{2\pi x} -1}~
\ln[{\cal Z}(ix+3/2)] - {\cal C.C.} -~\frac 1 2 \ln[{\cal Z}(3/2)]~,
\eeq
\beq
\Delta_4 \equiv
- {73\over 1536}\sinh^2\eta_0  -{251\over 3072}\sinh^4\eta_0
+{7\over 64}\sigma^2\sinh^2\eta_0 +{31\over 192}\sigma^2\sinh^4\eta_0 
-{1\over48}\sigma^4\sinh^4\eta_0~, 
\eeq
where
\beq
\phi(x) \equiv {1\over 3}\left(x-{1\over3}x^3\right)\ln {\cal Z}(1/x)~,
\eeq
\beq
{\cal Z}(x)\equiv 2^x \Gamma(1+x) P^{-x}_{\sigma-1/2}(\cosh\eta_0)~.
\eeq
Formulas (\ref{reszetazero}) and (\ref{reszetaprimezero}) represent the main result of our paper.

\section{Limiting cases}

The results obtained in the previous section for the zeta function and its derivative are valid
for a scalar field of arbitrary mass $m$ and coupling $\xi$. Here we focus on the specific case of
a massless conformally coupled field, as this allows us to compare our result to that of
Refs.~\cite{NS,MNSS}. By setting $\sigma=1/2$ in Eqs.~(\ref{reszetazero}) and
(\ref{reszetaprimezero}), the following expressions are found:
\bea
\zeta(0) &=& {1\over 6}{17\over 1920}~,\nn
\zeta'(0) &=& {1\over 6}{17\over 1920} \ln(\ell\sinh\eta_0)^2 
- {31\over 1536}\sinh^2\eta_0 - {131\over 3072}\sinh^4\eta_0 +\Delta_1~.
\label{ourcc}
\eea
These values can be compared with those of Ref.~\cite{MNSS}. After sorting out
some transcription errors in Tables I and II of the mentioned reference, the result is found to disagree by a
constant number. The question arises as to whether or not this difference is at all
significant. Clearly, this constant difference can be reabsorbed by redefining the renormalization
scale, and therefore it does not have any physical significance. However, from the mathematical
point of view, the origin of such a difference is not so clear-cut.

This disagreement has led us to consider another limiting case, 
which is obtained when the AdS curvature radius $\ell$ is large. This should reproduce
the ball geometry, and corresponds, in our terminology, to $\ell >> r_0$, i.e. $\eta_0<<1$.
This result has been computed by various authors using different techniques
\cite{BEK,BKD,Klausbook,BEGK,Dowker}, and therefore it should provide quite a robust check on the
result,
and also an alternative derivation, although more involved than necessary, of the zeta determinant
for the ball. Now, in the limit of $\eta_0 << 1$, we find
\bea
\zeta(0) &=& {1\over 6}{17\over 1920}~,\nn
\zeta'(0) &=& {1\over 6}{17\over 1920} \ln r_0^2 +{47\over 9216} -{1\over
64}\zeta'(0,1/2)\\
&+&{1\over 24} \zeta'(-1,1/2)+{1\over24}
\zeta'(-2,1/2)-{1\over 6} \zeta'(-3,1/2)+{1\over 12} \zeta'(-4,1/2)\nonumber~. 
\eea
This result is found in full agreement with those of Refs.~\cite{Klausbook,BEGK,Dowker},
thus providing a robust check of our result. We note that such a limiting case is not
recovered neither by the result of Refs.~\cite{NS,MNSS} nor by that of
Refs.~\cite{NOZ,ENOO}.

In order to find some explanation for the difference, let us briefly reconsider the method used
there. The first step of the method used in Refs.~\cite{NOZ,NS,ENOO,MNSS} is a conformal
transformation that changes the original background spacetime into a different one, where the 
evaluation of the zeta determinant is, in principle, easier. In Refs.~\cite{NOZ,NS,ENOO,MNSS} this
corresponds to a coordinate transformation, $dz = dr/a(r)$, upon which the original
line element becomes
\beq
ds^2 = a^2(r)(dz^2+ds^2_\Sigma)~.
\eeq
In other words, a conformal rescaling of the metric by $a(r)$ leaves us with a flat cylinder with
a de Sitter cross-section. This is the starting point taken in the above mentioned articles.
One immediately sees that the coordinate $r$ lies in the range of $0 \leq r \leq r_0$ in the
original frame, whereas in the conformally transformed one $z$ lies between $z(r_0) \leq z \leq
\infty$, implying that the conformally transformed spacetime corresponds to a semi-infinite
cylinder and a more important point is that the conformal transformation is well defined at every
point of the AdS bulk, except from the center where it breaks down. This observation
raises the question as to whether or not this procedure is actually valid. Certainly it requires care.
Obviously, in the two brane setup this problem does not exists since $r$ is never $0$.

\section{Concluding remarks}

The present article was devoted to providing an alternative derivation of the zeta function and
of the functional determinant for a scalar field on a de Sitter brane background, which consists
of a higher dimensional AdS bulk spacetime bounded by a de Sitter section.
We considered the general case of a non-zero mass and coupling to the scalar curvature, thus
generalizing previous results limited to zero mass and conformal coupling.

For simplicity, we considered the case of a five dimensional bulk spacetime and Dirichlet
boundary conditions. However, the result can be extended, with no additional technical problems
and modulo some algebra, to other boundary conditions or higher dimensionalities. The choice of
five dimensions was also motivated by the possible relevance of our calculation for the bulk
inflaton model proposed in \cite{KKS,HS}.

One of the interesting points of the approach developed here lies in the fact that we do not make
use of conformal transformations as is done in Refs.~\cite{NOZ,NS,MNSS,ENOO} and whose results, 
in the limit of large AdS radius does not reproduce that of the ball given in Refs.~\cite{Klausbook,BEGK,Dowker} 

The basic tools of our calculation were a contour integral representation for the zeta function and
asymptotics of the eigenfunctions, for which we have followed 
Refs.~\cite{BKK,BKD,Klausbook}. In particular, the method devised in \cite{BKK} has
proven to be very powerful, and extremely useful in the case discussed here.

As a check on the calculation, we have considered the limiting case of massless, conformally coupled
fields, which was found to disagree with the result of \cite{NS,MNSS}. The difference amounts to a
constant, which is physically harmless and can be removed by redefining the renormalization
scale.
This difference motivated us to consider another limiting case, which is obtained when the
curvature radius of the higher dimensional AdS space is very large, leading to a ball-like geometry, for which extensive calculations of the zeta
determinant are available \cite{Klausbook,BEGK,Dowker}. In such a limit we recover those results
and this should provide a robust check of our calculation, 
and, interestingly, this limit provides an alternative derivation, although technically
unnecessary, of the ball determinant.
We note that the results for the derivative of the zeta function given in Refs.~\cite{NOZ,NS,MNSS,ENOO} do not recover this limit. 

Various generalizations of the work presented here are possible. Extensions to higher
dimensionalities, different boundary conditions, two-brane setups and higher spin fields should follow without
additional difficulties and only a larger amount of algebra might be needed. Looking at the
possible relevance of these kinds of calculations in the bulk inflaton model and more generically
in brane world cosmology also deserves further study. We hope to report on these issues in our
future work.

\acknowledgments

We would like to thank, Y. Himemoto, S. Nojiri, S. Ogushi, O. Pujol{\`a}s, T. Tanaka and D. Toms
for useful comments. We thank I. Moss for poyinting out the ill-defined nature of the conformal transformation used in
Refs.~\cite{ENOO,MNSS}.
We also thank I. Moss, J. Norman and W. Santiago-Germ{\'a}n, for help in sorting out the
transcription errors in Ref.~\cite{MNSS}. A.F. is indebted to L. Da Rold, R. Fiore, J. Garriga and A. Papa 
for discussions at the early stage of this work and acknowledges the kind hospitality of
the Department of Physics of the Universit\`a della Calabria, where part of this
work was carried out. A.F. is supported by the JSPS under contract  No. P04724. W.N. is supported
by the Fellowship program, Grant-in-Aid for the 21COE, `Center for Diversity and Universality in Physics' at Kyoto University. M.S. is
supported in part by Monbukagakusho Grant-in-Aid for Scientific Research(S) No. 14102004. 

\appendix
\section{Uniform asymptotic expansion of the Legendre functions}
\label{uniasy}

\noindent 

Following Refs.~\cite{Khus} and \cite{Thorne}, the uniform asymptotic expansion of $P^{-\nu}_{\mu-1/2}(\cosh \eta)$ can be obtained. In this
appendix, we simply outline the procedure and extend the results to the case of the logarithm of the Legendre function $P_{\mu}^{\nu}(\cosh
\eta)$.

For large values of $\mu$, the solution of the Legendre differential equation can be written in the well-known WKB form. When substituted into
the original equation, this leads to a set of recursive equations that allow one to determine the expansion. Omitting the details, the result
is
\beq
P_{\mu-1/2}^{\nu}(\cosh \eta) = \sqrt{{t \over 2\pi\nu}}e^{\nu \xi}\mu^{-\nu}+ \sum_i \alpha_i \nu^{-i}~,
\eeq
where the functions $t$ and $\xi$ are given by
\beq
  t = {1\over\sqrt{1+\tau^2\sinh^2 \eta}}~, 
\label{t}
\eeq
and
\beq
\xi = \ln \frac{\tau\sinh \eta}{\sqrt{1+\tau^2\sinh^2 \eta}+\cosh \eta} - \tau\left[\tanh^{-1}{1\over\tau}-\tanh^{-1}{1\over\tau t\cosh
\eta}\right] + 1~.
\label{si}
\eeq
We have defined $\tau = \mu/\nu$ and the coefficients $\alpha_i$ can be computed recursively. They are quite lengthy for increasing values of
$i$ and since they are not used directly we do not report them. The interested reader is addressed to the work of Refs.~\cite{Thorne,Khus},
where they are also derived. 

It is a relatively simple task to get the logarithm of the previous expression and this can be achieved by using a symbolic manipulation program. 
The result is
\beq
\ln P^{-\nu}_{\mu-1/2}(\cosh \eta) \sim \ln\left[\sqrt{t\over2\pi\nu}\,e^{\nu\xi}\mu^{-\nu}\right] + \sum_{n=1}^{\infty}
{E_n(t,\tau)\over\nu^n}~, 
\label{asym}
\eeq
where it is essential to note that the coefficients $E_n(t,\tau)$ are bounded in the full range $0\leq\tau<\infty$ and for large $\tau$ they
scale as inverse powers of $\tau$. For $\nu\rightarrow\infty$ they exhibit a power law growth of finite order in $\nu$. This can be checked
for the first four coefficients $E_n$, which are found to be
\bea
  E_1 &=& -{1\over24(\tau^2-1)}\left(5\tau^2t^3\cosh^3 \eta-3t\cosh\eta+1-3\tau^2t\cosh \eta
\right), \nn
  E_2 &=& {1\over16(\tau^2-1)^2}\left(-1+t^2\cosh^2 \eta+7\tau^2t^2\cosh^2 \eta
-\tau^2-6\tau^4t^4\cosh^4 \eta+\tau^4t^2\cosh^2 \eta-6\tau^2t^4\cosh^4 \eta\right. \nn
       && \left.+5\tau^4t^6\cosh^6 \eta\right), \nn
  E_3 &=& -{1\over40320(\tau^2-1)^3}\left(118881\tau^4t^5\cosh^5 \eta
-51765\tau^2t^3\cosh^3 \eta+18270\tau^2t\cosh \eta-51765\tau^4t^3\cosh^3 \eta\right. \nn
       && \left.+2835\tau^4t\cosh \eta+33453\tau^6t^5\cosh^5 \eta
-2625\tau^6t^3\cosh^3 \eta-69615\tau^6t^7\cosh^7 \eta+33453\tau^2t^5\cosh^5 \eta\right. \nn
       && \left.-69615\tau^4t^7\cosh^7 \eta+38675\tau^6t^9\cosh^9 \eta-98+2835t\cosh \eta-294 \tau^2 
-2625t^3\cosh^3 \eta\right), \nn
  E_4 &=& {1\over128(\tau^2-1)^4}\left(565\tau^8t^{12}\cosh^{12} \eta+
1503\tau^4t^4\cosh^4 \eta-2282\tau^6t^6\cosh^6 \eta-288\tau^4t^2\cosh^2 \eta\right. \nn
       && \left.-1356\tau^6t^{10}\cosh^{10} \eta-288\tau^2t^2\cosh^2 \eta+
542\tau^2t^4\cosh^4 \eta-2282\tau^4t^6\cosh^6 \eta+542\tau^6t^4\cosh^4 \eta\right. \nn
       && \left.-18\tau^6t^2\cosh^2 \eta-284\tau^2t^6\cosh^6 \eta-284\tau^8t^6\cosh^6 \eta+
13\tau^8t^4\cosh^4 \eta+1062\tau^8t^8\cosh^8 \eta\right. \nn
       && \left.-1356\tau^8t^{10}\cosh^{10} \eta+5+30\tau^2+5\tau^4+13t^4\cosh^4 \eta-
18t^2\cosh^2 \eta+1062\tau^4t^8\cosh^8 \eta\right. \nn
       && \left.+3114\tau^6t^8\cosh^8 \eta\right). \label{ln6}
\eea

By inspecting the previous expansion and recalling the above-mentioned properties of the coefficients $E_k(t,\tau)$, one notices that it has
the same structure as Eq.~(\ref{phix}).


\section{Evaluation of $I^{\rm R}(z^2)$ and $I^{\rm pole}(z^2)$.}
\label{appprop}

\subsection{Evaluation of $[I^{\rm R}]_{\rm log}$.}

From Eq.~(\ref{asym}), we can find the various pieces that appear in Eqs.~(\ref{zetz}) and (\ref{zetp}). Let us first consider $[I^{\rm
R}]_{\rm log}$. From Eq.~(\ref{asym}), we see that the coefficient of the logarithmic piece, for large $z$, is 
\beq
  [I^{\rm R}]_{\rm log}=-\frac 1 2 \sum_{j} {d_j\over \nu_j^{2s}} \left( \nu_j+\frac 1 2\right)~.
\label{hurwitz}
\eeq
To deal with this sum, we introduce a generalized $\zeta$-function:
\beq
\zeta_\nu(s)=\sum_{j} d_j \nu_j^{-s}
\eeq
which can be expressed, in five dimensions, in terms of Hurwitz $\zeta$-functions. Trivial manipulations give
\beq
[I^{\rm R}]_{\rm log}=-\frac 1 2 \zeta_\nu(2s-1)-\frac 1 4 \zeta_\nu(2s)~,
\eeq
which, in five dimensions, becomes
\beq
[I^{\rm R}]_{\rm log}={1\over6}{17\over1920}  ~.
\label{Ilog}
\eeq

\subsection{Evaluation of $I^{\rm pole}(z^2)$.}

The term $I^{\rm pole}(z^2)$ is more involved to evaluate. We need to rewrite the uniform asymptotic expansion (\ref{asym}) in terms of
inverse powers of $\nu$ and then perform the summation over $j$. This will allow us to express, in five dimensions, expansion (\ref{asym}) in
terms of Hurwitz $\zeta$-functions from which the pole part can be extracted. The calculation is rather lengthy, although
straightforward. Here, we simply quote the result, which can be written as follows:
\beq
I^{\rm pole}(z^2) =  {1\over 6} \left[\omega_1(x) -{1\over 4}\omega_2(x) + \omega_3(x) + {1\over 8} \omega_4(x) \right]_{x=\ell z}~, \label{poleM}
\eeq
where we have defined the following quantities for notational convenience:
\bea
\omega_1 (x) &=& 
-{\sigma^2 \sinh^2\eta_0\over 16(1+x^2\sinh^2\eta_0)^4}
\left(4+(7-10x^2)\sinh^2\eta_0 +x^2(-8+x^2)\sinh^4\eta_0\right)~,\nn
\omega_2(x) &=& 
{\sinh^2\eta_0\over 16(1+x^2\sinh^2\eta_0)^3}\left(1-4x^2+x^2(-4+x^2)\sinh^2\eta_0\right)~,\nn
\omega_3(x) &=& 
{1 \over 128 (1+x^2\sinh^2\eta_0)^6}
\left[\right. \sinh^2\eta_0 \left(\right.8-32x^2+ \sinh^2\eta_0 (13-244x^2+288x^4\nn
&-&2x^2(116-313x^2+116x^4)\sinh^2\eta_0+x^4(288-244x^2+13x^4)\sinh^4\eta_0\nn
&+&8x^6(-4+x^2)\sinh^6\eta_0\left.\right)\left.\right]~,
\nn
\omega_4 (x)&=& 
{\sigma^2 \sinh^2 \eta_0 \over 1+x^2 \sinh^2\eta_0}
\left({1\over 2}+ {\sigma^2 \sinh^2 \eta_0 \over 1+x^2 \sinh^2\eta_0} \right)~.
\label{omegas}
\eea

The absence of logarithmic terms in $z^2$ implies that $[I^{\rm pole}]_{\rm log}=0$.
From the expression (\ref{poleM}), one easily finds that
\beq
[I^{\rm pole}]_{\rm reg} = \lim_{z^2 \rightarrow \infty} I^{\rm pole} (z^2) = 0
\label{ipreg}
\eeq
and
\beq
I^{\rm pole}(0)
=-\frac{\sinh^2\eta_0}{32}  \left(1+2\sinh^2\eta_0\right) \left(\sigma^2-1/4\right)
+ \frac{\,{\sinh \eta_0 }^4}{48}\left(\sigma^2-1/4\right)^2~,
\label{ipzu}
\eeq
both of which are required for the evaluation of $\zeta(0)$.

The integral of the pole piece is readily evaluated and the result we find is
\bea
\int^\infty_0 dz^2 \ln z^2 {d I^{\rm pole}(z^2)\over d z^2} =
&-&{47\over 9216}
+{73\over 1536}\sinh^2\eta_0 
+{251\over 3072}\sinh^4\eta_0 \nn
&-&{7\over 64}\sigma^2\sinh^2\eta_0  
-{31\over 192}\sigma^2\sinh^4\eta_0 
-{1\over 48}\sigma^4\sinh^4\eta_0 \nn
&+&{1\over 768}\sinh^2\eta_0 (4\sigma^2-1)(-6+\sinh^2\eta_0 (-13+4\sigma^2))\ln( \ell\sinh\eta_0)^2~. 
\label{logM}
\eea

\subsection{Evaluation of $I^{\rm R}(0)$.}

To evaluate $I^{\rm R}(0)$, we follow Ref.~\cite{BKK} once again and employ a more expedient approach based on the Abel-Plana summation
formula. For $z=0$ we have 
\bea
  \label{im_0}
  I(0,s)&=&\sum_j {f(\nu_j)\over \nu_j^{2s}}~,\\
  f(\nu_j)&=&\frac 1 3 \left(\nu_j^3-\frac 1 4 \nu_j\right) \ln[~2^{\nu_j}\Gamma(1+\nu_j)
~P^{-\nu_j}_{\sigma-\frac 1 2}(\cosh\eta_0)]~.
\eea
The validity of the form of the function, $f(\nu_j)$, which we use in the Abel-Plana summation formula is discussed in \cite{BKK}.
By applying the Abel-Plana formula, which allows one to convert the sum (\ref{im_0}) into an integral, we get
\beq
  I(0,s) = \int_{0}^\infty {f(\nu_x)\over \nu_x^{2s}}~dx+i\int_0^\infty {f(\nu_{ix})-f(\nu_{-ix})\over e^{2\pi x} -1}~dx+\frac 1 2 ~f(\nu_0)~, \label{summ}
\eeq
where we retain the regularizing factor, $\nu_j^{2s}$, only in the first term, because all other terms are finite as $s \rightarrow 0$. The only non-trivial term to compute in the expression is the first one. In order to evaluate it, we split the integral into three pieces, for convenience. Some algebraic manipulations give
\beq
\int_{0}^\infty {f(\nu_x)\over \nu_x^{2s}}~dx
= 
\int_{0}^1 x^{2s-6} \phi(x) dx 
+ \int_{1}^\infty x^{-6} \phi(x) dx 
- \int_{2/3}^\infty x^{-6} \phi(x) dx~,
\eeq
where we have put $s=0$ where possible. The function $\phi(x)$ is given by
\beq
\phi(x) = 
\frac 1 3 \left(x-\frac 1 4 x^3 \right) \ln\left[~2^{1/x}\Gamma(1+1/x)P^{-1/x}_{\sigma-\frac 1 2}(\cosh\eta_0)\right].
\eeq
Integrating the first two pieces by parts six times, we obtain
\bea
\int_{0}^\infty {f(\nu_x)\over \nu_x^{2s}}~dx &=&
\left[{x^{2s-5}\over 2s-5}\phi(x) - {x^{2s-4}\over (2s-5)(2s-4)}{d \phi(x)\over dx} + \cdots -
{x^{2s}\over (2s-5)(2s-4) \cdots 2s}{d^5 \phi(x)\over dx^5}\right]_{0}^{1} \nn
&+& \int_{0}^{1} {x^{2s} \over (2s-5)(2s-4) \cdots 2s}{d^6 \phi(x)\over dx^6} dx + 
\left[{x^{-5}\over -5}\phi(x) - {x^{-4}\over (-5)(-4)}{d \phi(x)\over dx} + \cdots -
{\ln x \over -120}{d^5 \phi(x)\over dx^5}\right]_{1}^{\infty} \nn
&-&{1\over 120} \int_{1}^{\infty} \ln x {d^6 \phi(x)\over dx^6} dx 
- \int_{2/3}^\infty x^{-6} \phi(x) dx~.
\eea
This can be appropriately analytically continued to $s=0$, giving
\beq
\int_{0}^\infty {f(\nu_x)\over \nu_x^{2s}}~dx =
{1\over 240s} {d^5 \phi(x)\over dx^5}|_{x=0}
+{137 \over 7200} {d^5 \phi(x)\over dx^5}|_{x=0} 
- \int_{0}^{\infty} \ln x {d^6 \phi(x)\over dx^6} dx 
- \int_{2/3}^\infty x^{-6} \phi(x) dx~.
\eeq
Summarizing, we find
\beq
\int_{0}^\infty {f(\nu_x)\over \nu_x^{2s}}~dx= {1\over s} I^{\rm pole}(0) + I_{*}^{\rm R}(0)
\eeq
where
\beq
  I^{\rm pole}(0) = {1\over240}\left.{d^5\phi\over dx^5}\right|_{x=0}~,
 \label{alternativeIpole}
\eeq
\bea
  I_{*}^{\rm R}(0) &=& {137\over7200}\left.{d^5\phi\over dx^5}\right|_{x=0} - {1\over120}\int_0^\infty dx\ln x{d^6\phi\over dx^6} - \int^{\infty}_{2/3} x^{-6} \phi(x)~dx~.
\label{gen_expbas}
\eea

A nice consistency check on the previous evaluation is given by the fact that the expression for 
$I^{\rm pole}(0)$, (\ref{ipzu}), evaluated previously using the asymptotic expansion agrees with (\ref{alternativeIpole}).

The previous results can be combined together to get $I^{\rm R}(0)$. We find
\bea
I^{\rm R}(0) &=&
{137\over7200}\left.{d^5\phi\over dx^5}\right|_{x=0} - {1\over120}\int_0^\infty dx\ln x{d^6\phi\over dx^6} 
- \int^{\infty}_{2/3} x^{-6} \phi(x)~dx \nn
&+&i\int_0^\infty dx{(ix+3/2)^3/3-(ix+3/2)/12\over e^{2\pi x} -1}~\ln[2^{ix+3/2} 
\Gamma(ix+5/2)P^{-ix-3/2}_{\sigma-1/2}(\cosh\eta_0)] + c.c. \nn
&&+\frac 1 2 \ln[2^{3/2} \Gamma(5/2)P^{-3/2}_{\sigma-1/2}(\cosh\eta_0)]~.
\label{iru}
\eea

\subsection{Evaluation of $[I^{\rm R}]_{\rm reg}$.}

We make the final effort to obtain the regular part of $I^{\rm R}(z^2)$. From previous arguments, we understand that such a piece comes from
the terms in the asymptotic expansion that scale as $z^0$, which then we have to sum over $j$. Thus, from \ref{asym}, we have, for
$s\rightarrow 0$, 
\beq
[I^{\rm R}]_{\rm reg} = -{1\over 2} (\ln2\pi + \ln (\ell \sinh \eta_0))\sum_j d_j + \sum_j d_j (\ln \nu_j +\ln \Gamma(\nu_j))~,
\label{sss}
\eeq
where we have used the fact that
\beq
\lim_{s\rightarrow 0} \sum_j d_j \nu_j^{1-2s} =0~.
\eeq
The first two terms can be computed easily, whereas to deal with the last sum in (\ref{sss})
we proceed as follows. First we use an integral representation for the logarithm of the $\Gamma-$function \cite{BEGK,Grad}, which gives
\beq
\sum_j d_j (\ln \nu_j +\ln \Gamma(\nu_j)) = {1\over 2}\ln 2\pi \sum_j d_j + \lambda + \theta~,
\eeq
where we have defined
\beq
\lambda \equiv  \sum_j d_j ({1\over 2}+\nu_j) \ln \nu_j~, 
\eeq
and
\beq
\theta \equiv \sum_j d_j \int_0^\infty 
\left({1\over 2}-{1\over t} + {1\over e^t-1}\right)
e^{-t\nu_j} {dt\over t}~.
\label{theta0}
\eeq
Let us first deal with $\theta$. It is possible to sum the series appearing in this expression by using the relation
\beq
\sum_j e^{-t\nu} = {e^{-t/2}\over e^t-1}~.
\label{expsum}
\eeq
Differentiating this relation one and three times, one immediately arrives at
\beq
\theta = {1\over 3}\int_0^\infty\left({1\over 2}-{1\over t} + {e^{-t}\over
1-e^{-t}}\right)\left[\left(-{d^3\over dt^3} {e^{-3t/2}\over 1-e^{-t}}\right) 
-{1\over 4}\left(-{d\over dt} {e^{-3t/2}\over 1-e^{-t}}\right)\right] {dt\over t}~,
\label{theta1}
\eeq
which can be expanded to write it in terms of elementary integrals:
\bea
\theta &=& \int_0^\infty\left[
-2{t^{\epsilon-2} e^{-9t/2} \over (1-e^{-t})^{4}}
-5{t^{\epsilon-2} e^{-7t/2} \over (1-e^{-t})^{3}}
-4{t^{\epsilon-2} e^{-5t/2} \over (1-e^{-t})^{2}}
- {t^{\epsilon-2} e^{-3t/2} \over (1-e^{-t})}\right.\nn
&&\left.+2{t^{\epsilon-1} e^{-11t/2} \over (1-e^{-t})^{5}}
+6{t^{\epsilon-1} e^{-9t/2} \over (1-e^{-t})^{4}}
+{13\over 2}{t^{\epsilon-1} e^{-7t/2} \over (1-e^{-t})^{3}}
+3{t^{\epsilon-1} e^{-5t/2} \over (1-e^{-t})^{2}}
+{1\over 2}{t^{\epsilon-1} e^{-3t/2} \over (1-e^{-t})}
\right] dt~.
\label{theta2}
\eea
Here we have introduced a regulating factor $\epsilon$, which ensures the convergence of the expression for $\Re \epsilon > 2 $. The limit
$\epsilon \rightarrow 0$ will be taken at the end.

All the above integrals can be evaluated starting from the standard formula
\beq
\varphi_1(a,b,\beta) \equiv \int_0^\infty {e^{-bt}\over 1-e^{-\beta t}} {dt\over
t^{-a}} = {1\over \beta^{1+a}} \Gamma(1+a) \zeta(1+a,b/\beta)~,
\eeq
which, by repeated differentiation with respect to $\beta$, produces the following relations
\beq
\varphi_2(a,b,\beta) \equiv \int_0^\infty {e^{-(b+\beta)t}\over (1-e^{-\beta t})^2}
{dt\over t^{-1-a}} = -{\partial\over \partial \beta} \varphi_1(a,b,\beta)~,
\eeq
\beq
\varphi_3(a,b,\beta) \equiv \int_0^\infty {e^{-(b+2\beta)t}\over (1-e^{-\beta
t})^3}
{dt\over t^{-2-a}} = 
-{1\over 2} \varphi_2(a+1,b,\beta)
-{1\over 2} {\partial\over \partial \beta} \varphi_2(a,b,\beta)~,
\eeq
\beq
\varphi_4(a,b,\beta) \equiv \int_0^\infty {e^{-(b+3\beta)t}\over (1-e^{-\beta t})^4}
{dt\over t^{-3-a}} = -{2\over 3} \varphi_3(a+1,b,\beta)
-{1\over 3} {\partial\over \partial \beta} \varphi_3(a,b,\beta)~,
\eeq
\beq
\varphi_5(a,b,\beta) \equiv \int_0^\infty {e^{-(b+4\beta)t}\over (1-e^{-\beta t})^5}
{dt\over t^{-4-a}} = 
-{3\over 4} \varphi_4(a+1,b,\beta)
-{1\over 4} {\partial\over \partial \beta} \varphi_4(a,b,\beta)~.
\eeq

The previous relation (\ref{theta2}) can then be expressed in terms of the functions $\varphi_j(a,b) \equiv \varphi_j(a,b,1)$. Simple
calculations lead to
\bea
\theta &=& 
-2 \varphi_{4}(\epsilon-5,3/2)
-5\varphi_{3}(\epsilon-4,3/2)
-4\varphi_{2}(\epsilon-3,3/2) 
-\varphi_{1}(\epsilon-2,3/2)\nn
&&+2\varphi_{5}(\epsilon-5,3/2)
+6\varphi_{4}(\epsilon-4,3/2)
+{13\over 2}\varphi_{3}(\epsilon-3,3/2)
+3\varphi_{2}(\epsilon-2,3/2)
+{1\over 2}\varphi_{1}(\epsilon-1,3/2)~.
\label{theta3}
\eea
The limit $\epsilon \rightarrow 0$ can now be taken and the result is found to be
\beq
\theta = -{1\over 64}\zeta'(0,1/2)-{1\over 24}\zeta'(-2,1/2)+{5\over 12}\zeta'(-4,1/2)~.
\eeq
The term $\lambda$ can be evaluated, by using the following relation:
\beq
{d\over ds} \sum_j \nu_j^{a-s} = -\sum_j \nu_j^{a-s}\ln \nu_j ~.
\eeq
A straightforward computation leads to
\beq
\lambda = -{1\over 6} \zeta'(-3,1/2)+{1\over 24} \zeta'(-1,1/2)-{1\over 3} \zeta'(-4,1/2)+{1\over
12} \zeta'(-2,1/2)~.
\eeq
Some simple algebra allows us to combine the previous results to arrive at
\beq
[I^{\rm R}]_{\rm reg} = {17\over 6}{1\over 960} \ln (\ell \sinh \eta_0)
-{1\over 64}\zeta'(0,1/2)+{1\over 24} \zeta'(-1,1/2)
+{1\over24} \zeta'(-2,1/2)-{1\over 6} \zeta'(-3,1/2)
+{1\over 12} \zeta'(-4,1/2)~.
\label{IReg}
\eeq


\end{document}